\documentclass[epj,final]{svjour}
\usepackage{epsfig,amssymb,bm,color}
\usepackage{epstopdf}
\usepackage{graphicx}
\usepackage{dcolumn}
\usepackage{amsmath}
\usepackage{bm}
\usepackage[english]{babel}

\begin{document}
\title{Assisted dynamical Schwinger effect: pair production in a pulsed bifrequent field}
\titlerunning{Assisted dynamical Schwinger effect in a pulsed bifrequent field}
\authorrunning{A.D. Panferov et al.}

\author{A.~D.~Panferov\inst{1}
\and
S.~A.~Smolyansky\inst{1}
\and
A.~Otto\inst{2,3}
\and
B.~K\"ampfer\inst{2,3}
\and
D.~B.~Blaschke\inst{4,5}
\and
{\L}.~Juchnowski\inst{4}
}
\institute{Department of Physics, Saratov State University, 410026 Saratov, Russia
\and
Institute of Radiation Physics, Helmholtz-Zentrum
Dresden-Rossendorf, 
01328 Dresden, Germany
\and
Institut f\"ur Theoretische Physik, Technische Universit\"at Dresden, 
01062 Dresden, Germany
\and
Institute for Theoretical Physics, University of Wroclaw, 50-204 Wroclaw, Poland
\and
Bogoliubov Laboratory for Theoretical Physics, Joint Institute for Nuclear Research, 141980 Dubna, Russia
}

\date{Received: \today / Revised version: date}

\abstract{
Electron-positron pair production by the superposition of two 
laser pulses with different frequencies and amplitudes is analyzed as a particular realization 
of the assisted dynamic Schwinger effect.  It is demonstrated that,
within a non-perturbative kinetic equation framework,
an amplification effect is conceivable for certain parameters. When both pulses have wavelengths longer than the Compton wavelength, the residual net density of produced pairs is determined by the resultant field strength. The number of pairs starts to grow rapidly if the wavelength of the high-frequency laser component gets close to the Compton wavelength.
}
\maketitle

\section{Introduction \label{sect:in}}

The possibility of direct energy conversion processes from a strong electromagnetic field into $e^-e^+$ pairs is one of the curious features of quantum electrodynamics (QED)
\cite{Sauter:1931zz,Heisenberg:1935qt,Schwinger:1951nm}.
However, the required critical electric field strength has the so-called 
Sauter-Schwinger value\footnote{We use $\hbar=c=k_B=1$ throughout this work.} 
$E_c \equiv m^2/ \vert e \vert=1.3\cdot 10^{16}$ V/cm
(here, $m$ and $e$ are the mass and the charge of the electron, resp.) 
which makes it inaccessible to direct experimental observations at present. 
The hope for the observation of such processes was revived with the advent of ultra-intensity laser systems in the optical or X-ray regimes \cite{Blaschke:2008wf}. 
The rapidly evolving laser technologies \cite{DiPiazza:2011tq}
triggered repeatedly the theoretical  search
for suitable laser configurations which have the potential to realize  pair production by Schwinger-type tunneling processes (for different variants, see \cite{exhilp}). 
A new avenue was provided by the dynamically assisted Schwinger effect 
\cite{Schutzhold:2008pz,Dunne:2009gi}, meaning that the tunneling path is abbreviated by an assisting second field, thus enhancing the originally small tunneling probability.  
Given this scenario, a number of dedicated investigations aimed at further
elaborating the prospects to find appropriate signals of the Schwinger effect.

Because of the important implications for related effects in other fields in physics
(see \cite{Dabrowski:2014ica,Ruffini:2009hg} 
for an overview including particle production in cosmology and astrophysics,
Hawking-Unruh radiation as well as conceptional issues of vacuum definition), 
many investigations address either the principles of the
strictly non-perturbative pair production \cite{Blaschke:2013ip} or employ special field models
to elucidate the general features, often only by numerical evaluation. 

The term "assisted Schwinger effect" stands for pair production from the vacuum under the influence of two fields - one assisting the other.
Special field models are, for instance, particular pulses (such as the Sauter- or the Gauss-pulse) 
or oscillating fields with particular envelopes (such as Sauter- or Gauss-pulse with sub-cycle structures). Since in a spatially homogeneous electric field the three-momentum of a charged particle is a good quantum number which makes the mode expansion appropriate, one often restricts oneself to such homogeneous fields. 
The rationale  for many models with a purely temporal dependence is that counter-propagating, suitably linearly polarized (laser) beams \cite{Piazza:2004sv} in the homogeneity region of anti-nodes represent such spatially constant fields. 
The account for spatial gradients is quite challenging \cite{Kleinert:2008sj,Schneider:2014mla}
and requires much more efforts.

The enhancement of Schwinger type pair production by an assisting field has been
con\-si\-de\-red already in \cite{Akal:2014eua,Li:2014psw,Nuriman:2012hn,Kohlfurst:2012rb,Linder:2015vta,Orthaber:2011cm,Hebenstreit:2014lra,Otto:2014ssa,Otto:2015gla}, e.g., for
\begin{itemize}
\item[(i)] a constant field plus some pulse with or without sub-cycle structures \cite{Linder:2015vta},
\item[(ii)] a superposition of two pulses without sub-cycles \cite{Orthaber:2011cm},
\item[(iii)] a superposition of two pulses with oscillating sub-cycle structure \cite{Hebenstreit:2014lra,Otto:2014ssa,Otto:2015gla}.
\end{itemize}
In the latter case, the common envelope was taken with a long flat-top period with
short ramping and de-ramping stages. 
Besides numerical examples, also the underlying enhancement mechanism has been clarified for that special field model:
It is the shift of the relevant zero of the quasi-particle energy in the complex time
domain toward the real axis (cf.~\cite{Linder:2015vta,Dumlu:2011rr} for other field configurations).
Here we are going to extend the considerations in Refs.~\cite{Otto:2014ssa,Otto:2015gla} and study,
by numerical means, some systematics of the enhancement for  a Gauss envelope.
Besides the oscillation frequencies of both fields, the temporal width of the
Gauss envelope enters as relevant new parameter related to time scales. 

Our paper is organized as follows. In section \ref{sect:description} we recall the formal framework of
the quantum kinetic equations as basis of our non-perturbative analysis.
In section \ref{sect:fields} we introduce the para\-metri\-za\-tion of the field model we consider.
Numerical results are presented in section \ref{sect:results}. 
In section \ref{sect:disco} we give a critical discussion of the explored parameter range w.r.t.\ applications, 
and in section \ref{sect:sum} we present the summary of this work.
      
\section{Theoretical basis  \label{sect:description}}

The non-perturbative consequence of the equations of motion of QED determines 
the vacuum effects in a given external, spatially homogeneous electric field with an arbitrary time dependence \cite{Grib:1994}.
For instance, one can employ the quantum kinetic equation \cite{Schmidt:1998vi}
describing the $e^- e^+$ creation by an electric field
$E(t) = - \partial_t A(t) \equiv -\dot{A}(t)$ with the four-vector potential in Hamilton gauge
(we use natural units with $c = \hbar = 1$), $A^\mu(t)=(0,0,0,A(t))$,
\begin{equation}\label{ke}
\dot w(\mathbf{p} ,t) = \frac{\lambda(\mathbf{p}, t )}{2}
\int\limits^t_{t_0} dt^{\prime} 
\lambda(\mathbf{p}, t^{\prime}) w(\mathbf{p}, t^{\prime})
\cos\theta(\mathbf{p}, t, t^{\prime}),
\end{equation}
where $w(\mathbf{p}, t)=1-2f(\mathbf{p}, t)$ is the depletion function containing the 
dimensionless phase space distribution function per spin projection degree of freedom 
$f(\mathbf{p}, t) = d N(\mathbf{p}, t) / d^3 p \, d^3 x$,
and 
\begin{equation}
 \lambda(\mathbf{p},t) =
\frac{e E(t) \, \varepsilon_{\bot} (p_\perp)}{\varepsilon^{2}(\mathbf{p}, t)}~, 
\label{lambda}
\end{equation}
is the amplitude of the vacuum transition, while
\begin{equation}
\theta(\mathbf{p}, t, t^{\prime}) = 2 \int^t_{t^{`}} d\tau \, \varepsilon (\mathbf{p}, \tau)
\label{phase}
 \end{equation}
stands for the dynamical phase, describing the vacuum oscillations modulated by the external field.
The quasiparticle energy $\varepsilon$, the transverse energy $\varepsilon_\bot$ 
and the longitudinal quasiparticle momentum $P$ are defined as
\begin{eqnarray}
\varepsilon(\mathbf{p}, t) &=& \sqrt{\varepsilon^2 (p_\perp) + P^2(p_\parallel, t)} ,
\label{eq:energy}
\\
\varepsilon_\perp (p_\perp)&=& \sqrt{m^2 + p^2_\bot},
\label{eq:energy_perp}
\\
 P(p_\parallel, t) &=& p_\parallel -eA(t),
\label{eq:p-long}
\end{eqnarray}
where $p_\bot=|\mathbf{p_\bot}|$ is the modulus of the momentum component 
perpendicular to the electric field, and $p_\parallel$ stands for the
momentum component parallel to $E$. 

The integro-differential equation (\ref{ke}) is useful for the low-density approximation
by setting $f(t^\prime) \to 0$. For the complete  numerical evaluations of  (\ref{ke})
an equivalent system of ordinary differential equations is comfortable
\begin{eqnarray}
\label{ode-u}
\dot{u} & = &  - 2 \varepsilon v  + \lambda \sqrt{1 - u^2 - v^2}, \\
 \dot{v} & = & + 2 \varepsilon u~, 
\label{ode-v}
\end{eqnarray}
with $u$ and $v$ as auxiliary functions  being related via $u^2 + v^2 + w^2 =1$. 
Since the modes with momenta $\mathbf{p}$ decouple we have suppressed these arguments here, 
as well as the time dependence of all quantities.
Sometimes, the relation $\dot f = \lambda u/2$ is useful for a field acting a finite time only,
telling that, since $E(t) \to 0$ implies $\lambda \to 0$, $f \to {\rm const}$ for $E(t) \to 0$.
Initial conditions at the
remote past, $t_0 \to - \infty$, are $u(t_0) = v(t_0) = f(t_0) = 0$. 

As emphasized, e.g., in \cite{Dabrowski:2014ica}, a sensible quantity is
$\lim \limits_{t \to \infty} f(\mathbf{p}, t)$, since
the adiabatic particle number per mode depends on the chosen basis. Accordingly,
the residual pair number density is
\begin{equation}\label{dens_r}
n =  \lim \limits_{t \to \infty} 2 \int \frac{d\mathbf{p}}{(2 \pi)^3} f(\mathbf{p},t).
\end{equation}
The factor two refers to the two spin degrees of freedom which are summed up
since in a purely electric field the spin degrees of freedom are degenerate.

Other formulations of the basic equations are conceivable, e.g., by relating $f$
to the reflection coefficient at (above) an effective potential, where the problem's
heart is a Riccati equation \cite{Linder:2015vta,Dumlu:2011rr}.  
In such a way the equivalence
with a quantum mechanical scattering problem is highlighted, \\
where the potential is related to $\varepsilon(\mathbf{p}, t)$.
This makes evident that the residual
phase space distribution can, in general, obey an intricate momentum dependence.

Asymptotic methods for the solution of the kinetic equation~\eqref{ke} were developed
in~\cite{smolyansky_laser_2013,smolyansky_smoothed_2014}.
There, some difficulties of applying such methods for field parameters corresponding to 
the case of tunneling regime are also discussed.

\section{Field models \label{sect:fields}}

Only for a few cases the equations of section \ref{sect:description} allow for exact solutions. Most notable
are the Schwinger field $E_\text{Schw} = {\rm const}$ and the Sauter pulse
$E_\text{Saut} \propto 1/ \cosh^2(t/\tau)$ with a time scale $\tau$.
For a systematic approach to relate features of the residual momentum distribution and the
temporal field shape, see \cite{Dumlu:2011rr}. 
Therefore, in most cases of interest, one has to resort to numerical solutions. 
Here one faces the problem that, for pulses with or without sub-cycle structures,
a number of parameters determine the solution which can sensitively (often 
non-linearly) depend on the location in parameter space. Therefore, suitable
approximations and estimates are very important. For instance, in a WKB type
analysis the locations of zeroes of $\varepsilon$ in the complex $t$ plane are 
identified as important quantities determining the dominating exponential
factor for the pair production. This also explains that pulses which look similar
on the real $t$ axis can have strikingly different implications since the
analytic properties can be rather distinctive. On a qualitative level, the enhanced
pair production in the assisted dynamical Schwinger effect can be traced back
to moving the relevant zeroes towards the real axis (cf.~\cite{Linder:2015vta}),
as mentioned above.

A subject of intense previous studies 
\cite{Hebenstreit:2009km,Hebenstreit:2011pm}
was the Gauss pulse with sub-cycle structure
or, equivalently, a periodic field with Gaussian envelope
\begin{eqnarray} 
E(t) &=& E_0  \cos{ (\omega t+\varphi) } \exp{(-\frac{t^2}{2\tau^2})}, \label{field_G}  \\
A(t) &=& -\sqrt{\frac{\pi}{8}} E_0\tau \exp{(-\frac12 \sigma^2+i\varphi)} \label{field_GA} 
\nonumber \\
&& \times\text{erf}\left(\frac{t}{\sqrt{2}\tau} -i\frac{\sigma}{\sqrt{2}}\right) 
+ c.c. , 
\end{eqnarray}
where $E_0$ is the amplitude, $\omega$ denotes the oscillation frequency and $\varphi$ is the
carrier envelope phase, which determines the symmetry properties w.r.t. time reversal. 
Hereafter, we put $\varphi = 0$.
The parameter $\sigma= \omega \tau$ characterizes the number of oscillations within the pulse. 
For $\sigma > 4$, the known
examples \cite{Hebenstreit:2011pm} exhibit $f(t \to \infty)$ at $p_\perp =0$ as a 
strongly oscillating (in tune with $\tau$) function of $p_\parallel$ around a bell-shaped mean, 
the latter one accessible via a WKB approximation. 
The occurrence of two time scales,
$1/\omega$ and $\tau$, allows to define two Keldysh parameters,
$\gamma_\omega = (\omega/m) (E_c /E_0)$ and
$\gamma_\tau = 1/(m \tau) (E_c/E_0)$. Usually, $\gamma_\omega \ll 1$ 
is attributed \cite{Brezin:1970xf} to the tunneling regime
and can be termed dynamical Schwinger effect.

Considering (\ref{field_G}),  (\ref{field_GA}) as the strong pulse in the spirit of the
assisted dynamical Schwinger effect, one adds a second weak assisting pulse with the same
envelope form but different parameters yielding an eight dimensional parameter space
for the two-dimensional $p_\perp -  p_\parallel$ distribution. 
Here, the optimization theory 
\cite{Kohlfurst:2012rb,Hebenstreit:2014lra} is certainly very useful to search for 
parameters suitable for maximum amplification. Upon restricting to a narrow patch
in the parameter space  one can constrain the ansatz for the superposition of a
strong and a weak pulse, each with sub-cycles, to     
\begin{eqnarray} 
E(t) &=&  
E_0  \left\{ \cos{ (\omega t) } + k_E  \cos{ (k_\omega \omega t }) \right\} 
{\rm e}^{-t^2/(2\tau^2)}, \nonumber \\
\label{field_12G} \\
A(t)  &=&  -\sqrt{\frac{\pi}{2}} E_0 \tau \label{field_12GA} \times \Big\{ \exp (-\frac{\tau^2\omega^2}{2})
\nonumber\\
&&\text{Re} \left[ \text{erf} \left( \frac{t}{\sqrt{2}\tau} +i \frac{\tau \omega}{\sqrt{2}}\right) \right] 
\nonumber\\
&&+ k_E \exp\left(-\frac{\tau^2(k_\omega \omega)^2}{2}\right)
\nonumber\\ 
&&
\text {Re}\left[ \text{erf} \left( \frac{t}{\sqrt{2}\tau} +i \frac{\tau k_\omega \omega}{\sqrt{2}}\right) \right] \Big\}. 
\end{eqnarray}
In these expressions, $k_E \le 1$ is the field strength fraction of the amplitude of the weak pulse,
and $k_{\omega} \ge 1$ is the frequency ratio.  The envelopes of both pulses are synchronized
and the carrier envelope phases are dropped, leading to a $t \to -t$ symmetric field $E(t)$.
Thus, we are going to quantify the assisted dynamical Schwinger effect for moderate 
values of $k_{E, \omega}$ and $\tau$ in the mildly sub-critical regime with
$E_0 < E_c$ and $\omega \le m$. Having more extreme conditions in mind, e.g.
$k_\omega \ggg 1$, another field model could be more suitable, such as 
\begin{equation}
E(t) = E_0 ( 1 + k_E \cos \omega_2 t) \times {\rm envelope} 
\label{new_pulse}
\end{equation}
and the related function $A(t)$.
Beyond the Gauss envelope, 
super-Gauss or Sauter shapes should be considered in separate work, as also the impact of the
nonzero carrier envelope phases.   

\begin{figure*}
\includegraphics[width=0.33\textwidth]{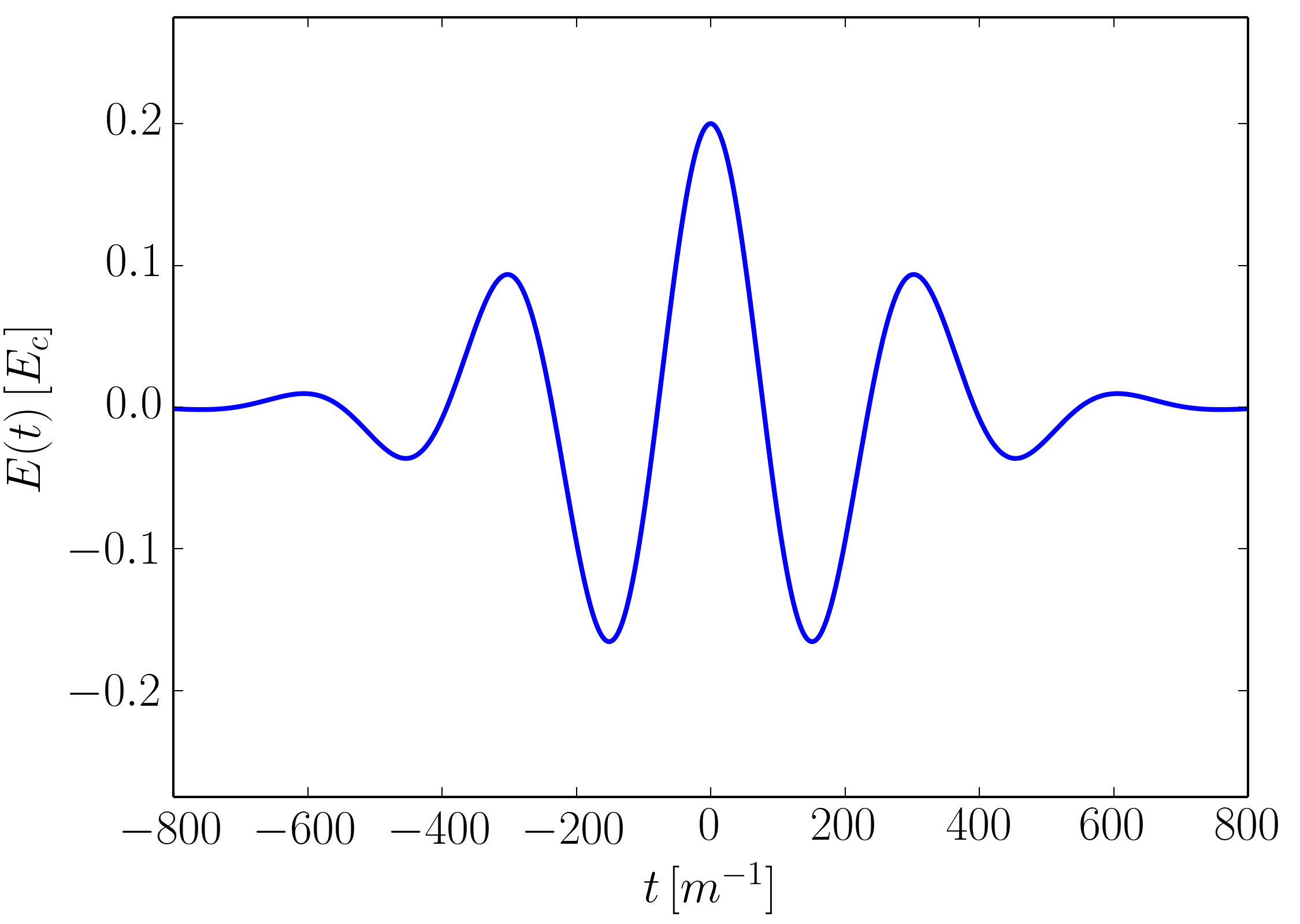}\hfill
\includegraphics[width=0.33\textwidth]{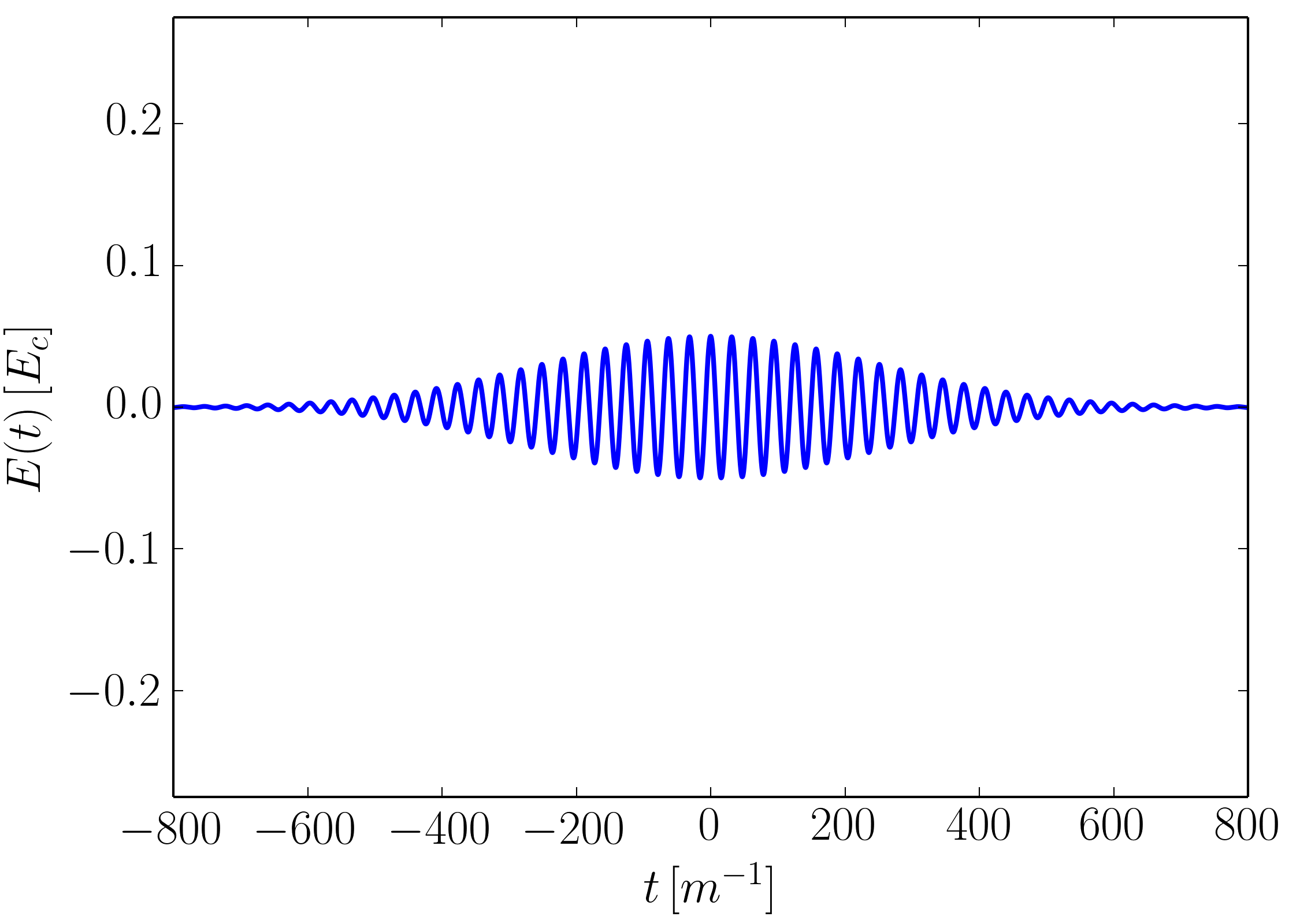}\hfill
\includegraphics[width=0.33\textwidth]{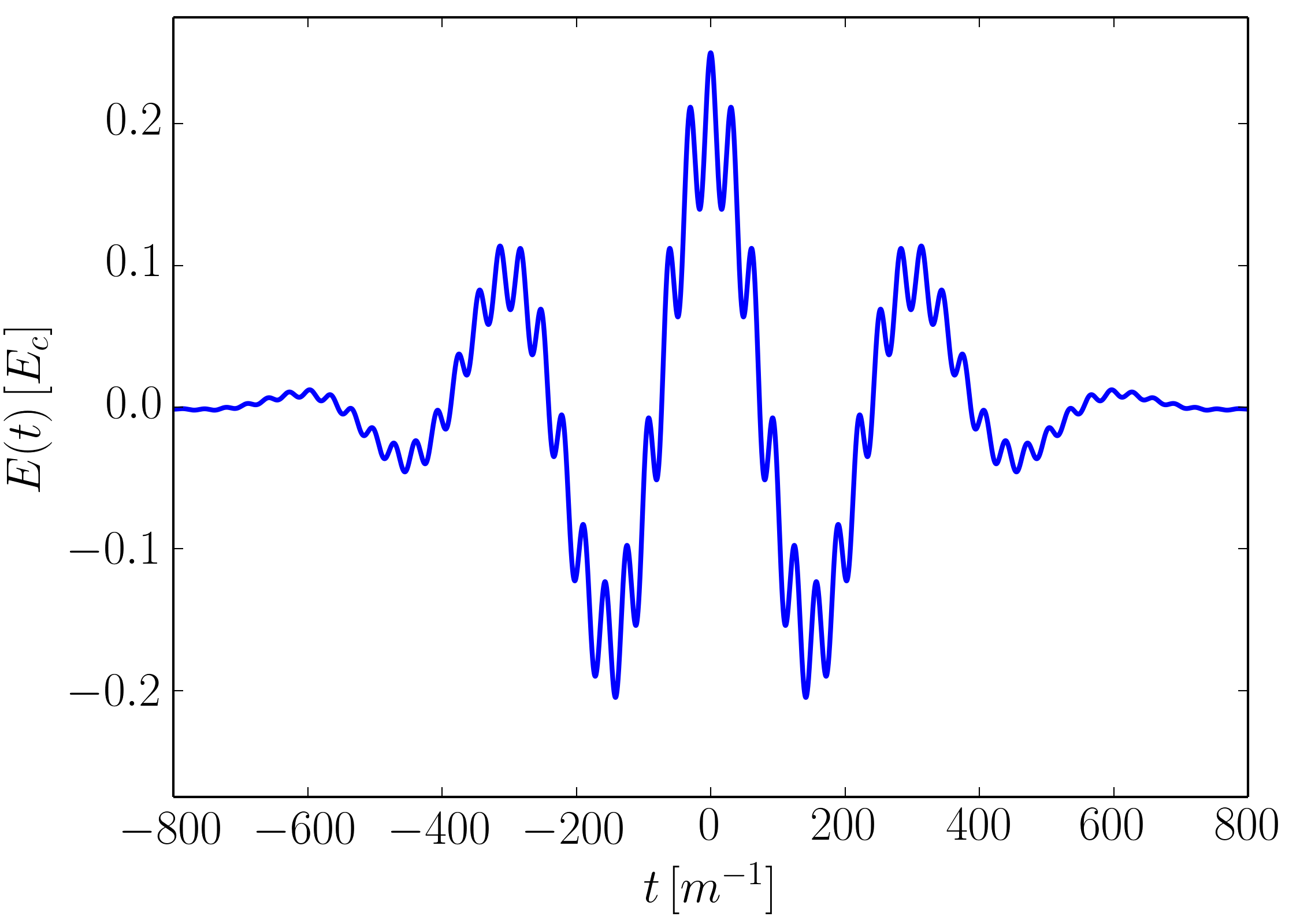}\\
\includegraphics[width=0.33\textwidth]{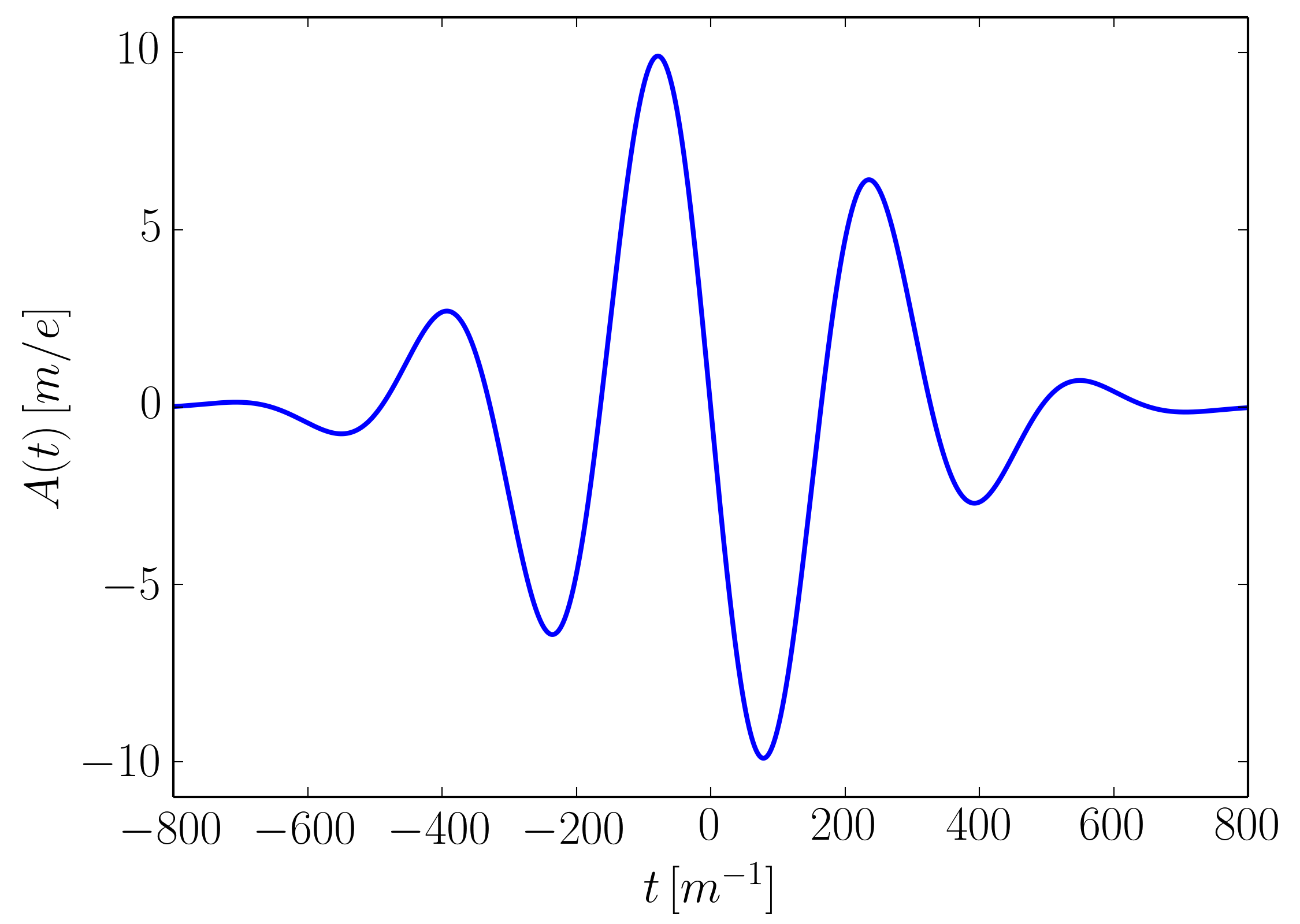}\hfill
\includegraphics[width=0.33\textwidth]{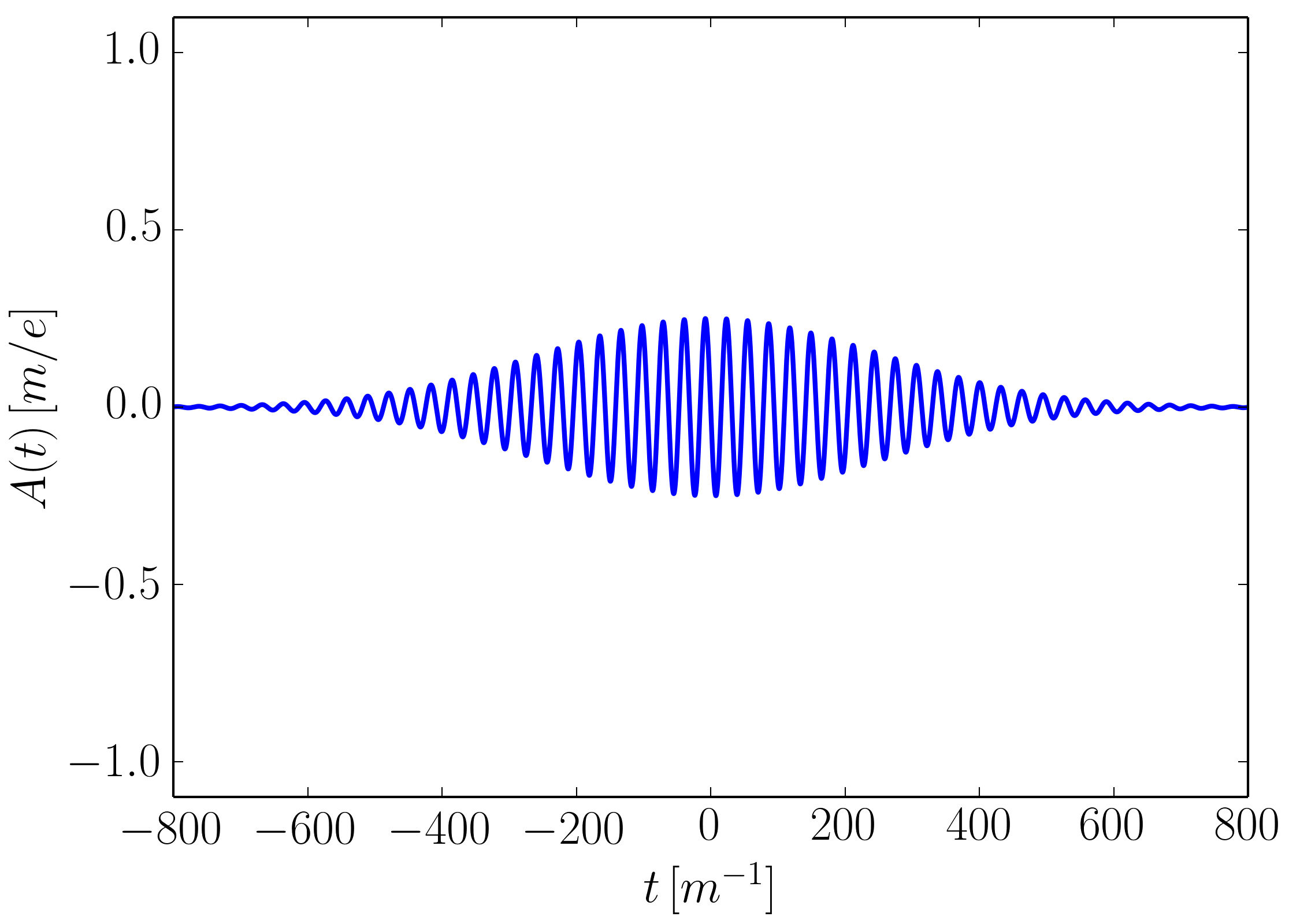}\hfill
\includegraphics[width=0.33\textwidth]{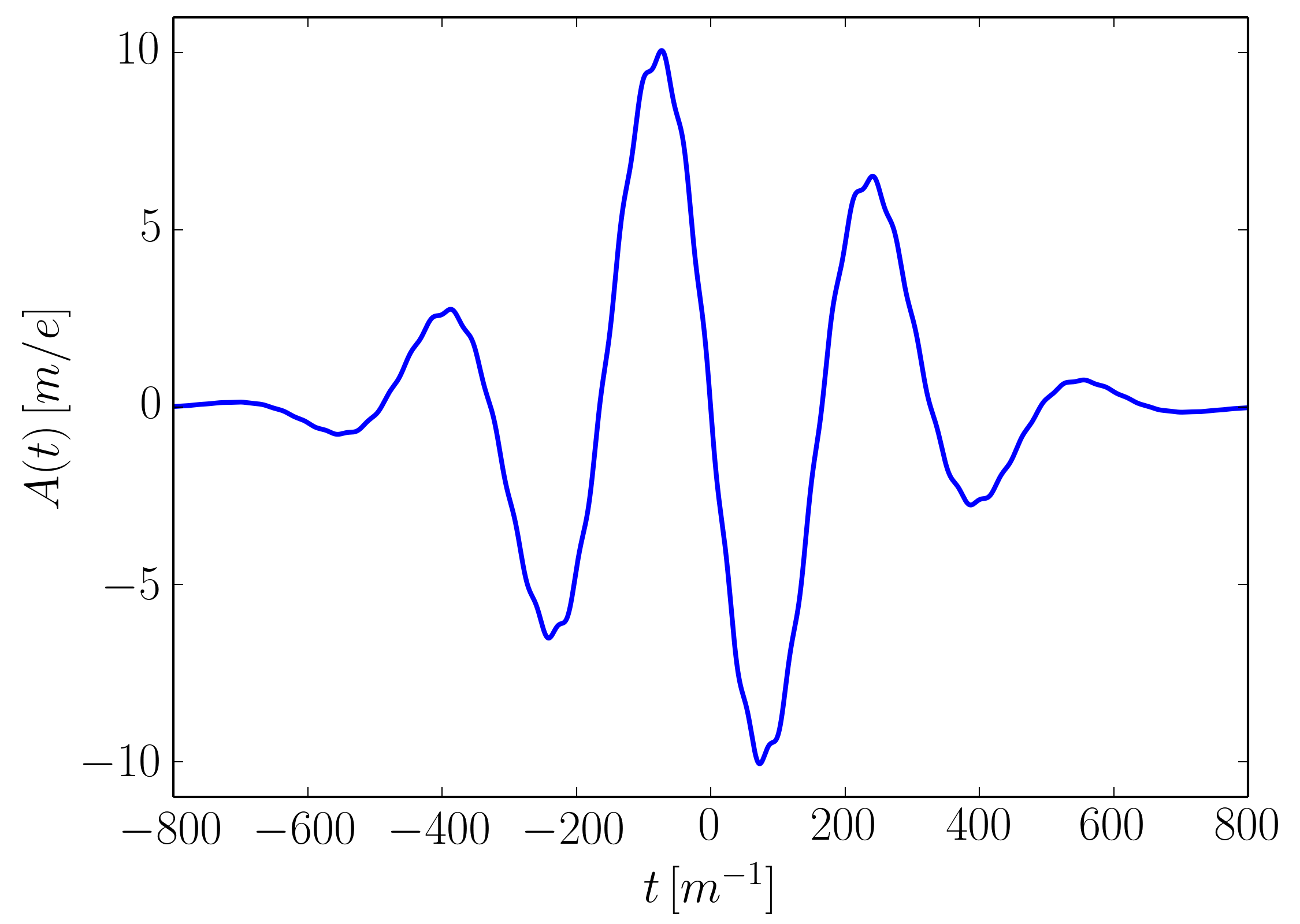}
\caption{The time dependence of the electric field (\ref{field_12G}) (upper row)
and the potential (\ref{field_12GA}) (lower row) for 
$E_0 = 0.2~E_c$, $\omega = 0.02~m$,
and $\tau = 5/\omega$.
Left column: the strong, low-frequency component "1" of the pulse,
i.e. only the first term in 
curly brackets in Eqs.~(\ref{field_12G}) and (\ref{field_12GA}) corresponding to $k_E = 0$;
middle column: the weak, high-frequency component "2" with $k_E = 0.25$ and 
$k_\omega = 10$, i.e. only the second term in 
curly brackets in Eqs.~(\ref{field_12G}) and (\ref{field_12GA});
right column: the superposition "1+2", i.e. the complete expressions in Eqs.~(\ref{field_12G}) and (\ref{field_12GA}).
\label{fig:1}}
\end{figure*}

Figure \ref{fig:1} shows an example of the electric field (upper row) and the potential (lower row)
of the strong, low-frequency pulse (left column, field "1" characterized by $E_0$, $\omega$, $\tau$
in (\ref{field_G}) and (\ref{field_GA})), the weak, high-frequency pulse 
(middle column, field "2" characterized by $k_E E_0$, $k_\omega\omega$, $\tau$ to be
used in (\ref{field_G}) and (\ref{field_GA}) instead of  $E_0$, $\omega$, $\tau$) 
and the superposition of both 
(right column, field "1+2" according to (\ref{field_12G}) and (\ref{field_12GA})).
We emphasize the much more pronounced ``roughening'' of the electric field ``1+2'' by ``2'',
while the impact on the potential looks very modest
(note the different scales of left and middle panels in the bottom row).

\section{Numerical results \label{sect:results}}

\begin{figure*}
\includegraphics[width=0.33\textwidth]{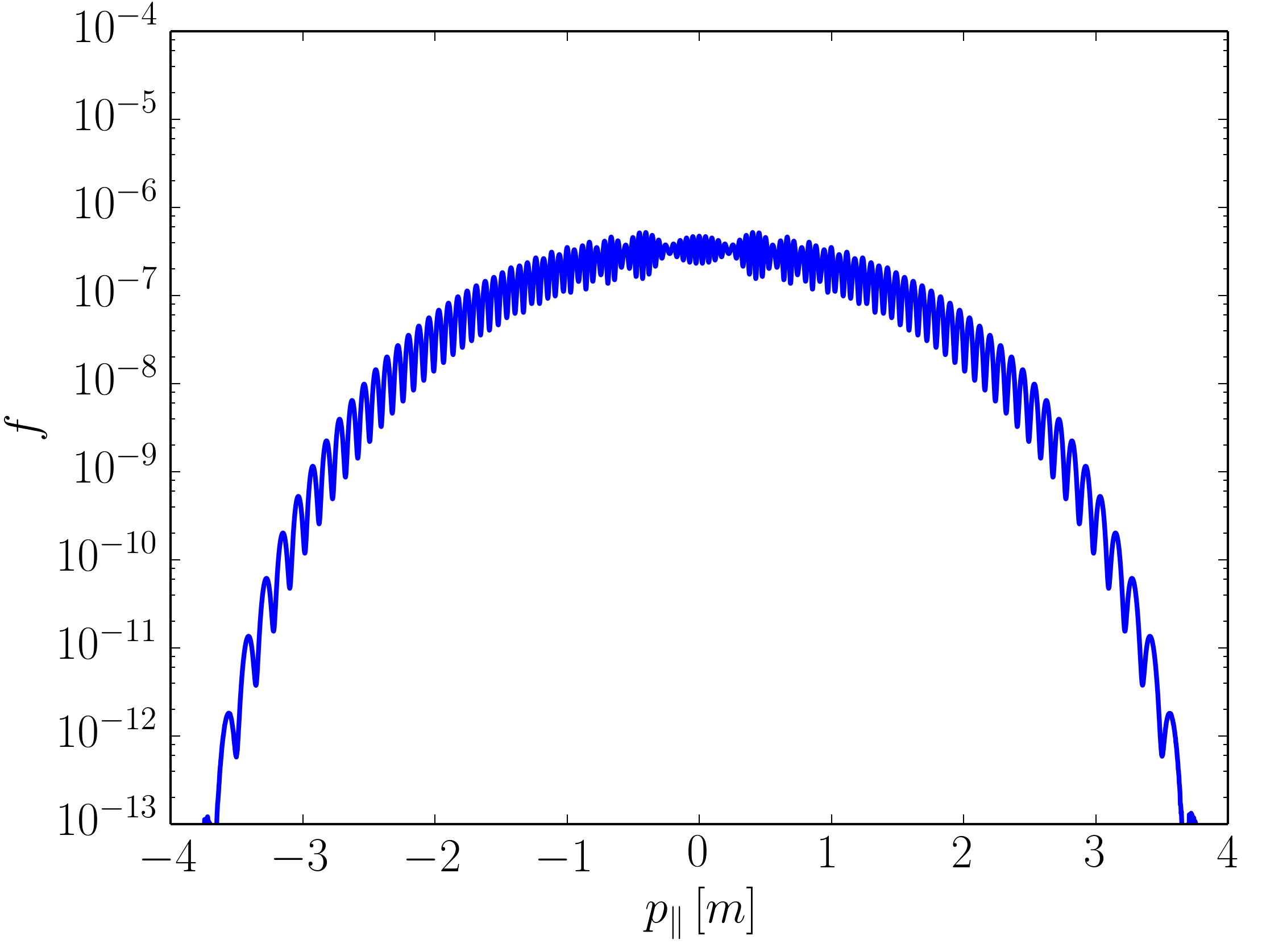}\hfill
\includegraphics[width=0.33\textwidth]{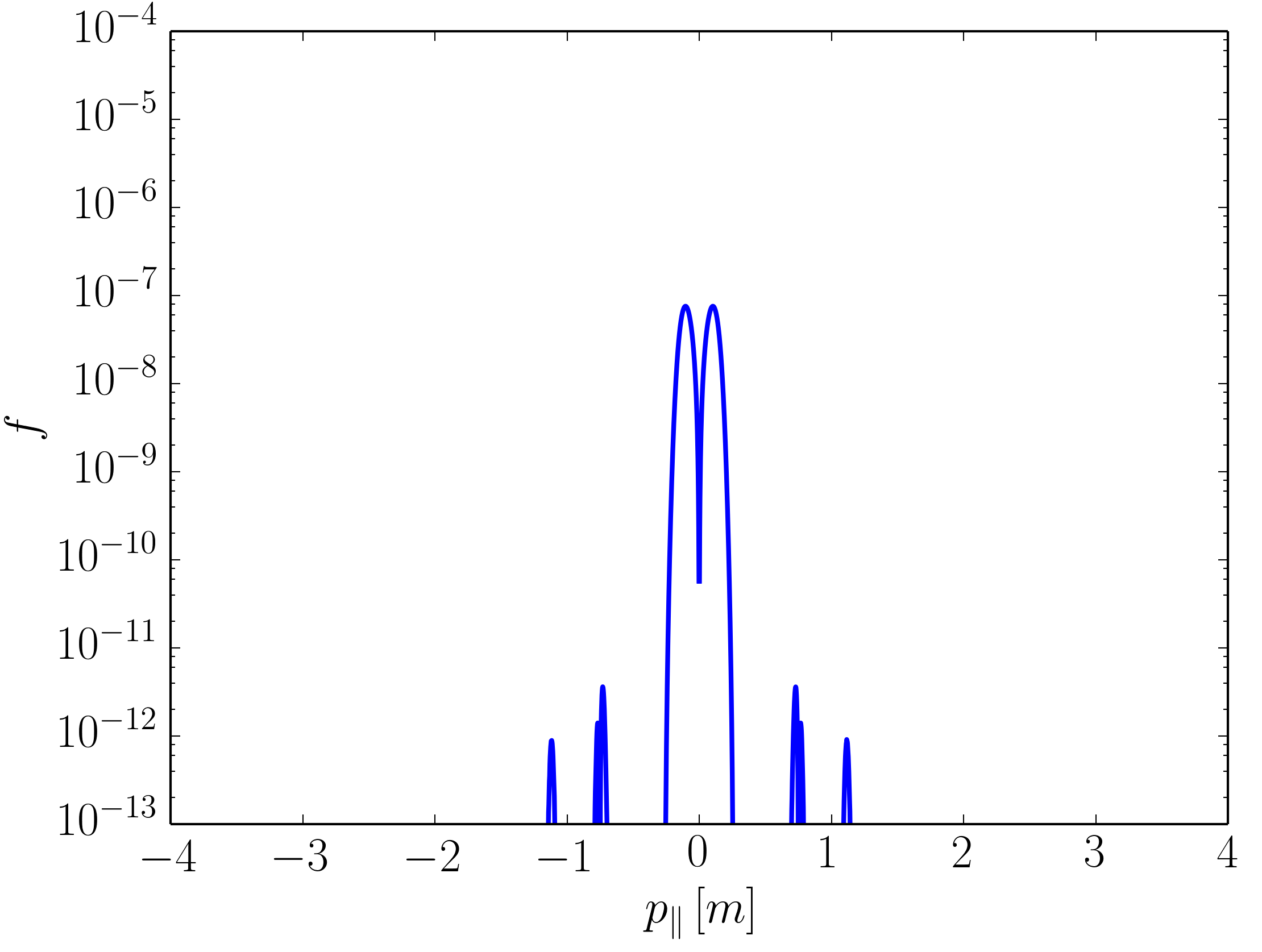}\hfill
\includegraphics[width=0.33\textwidth]{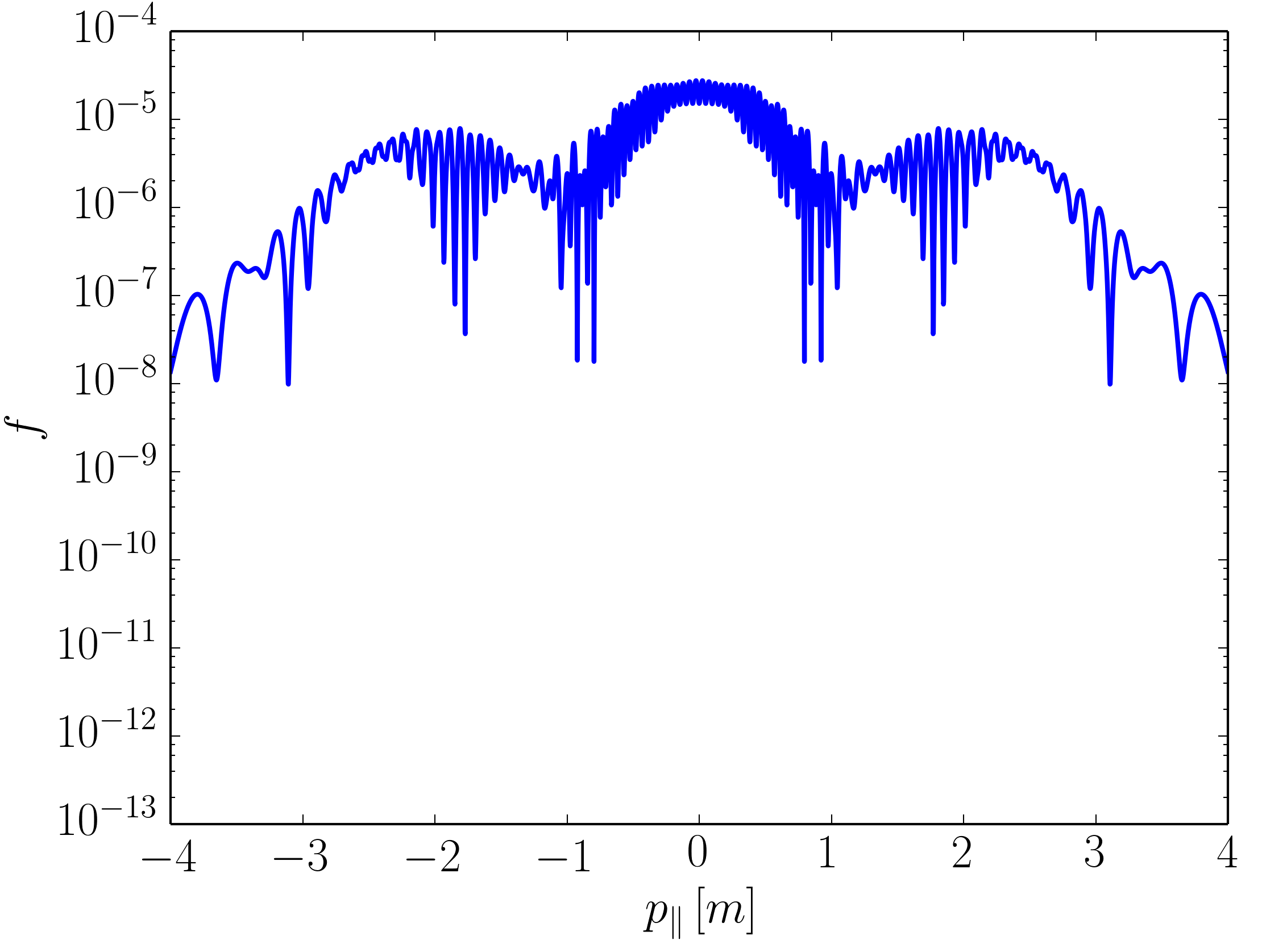}\\
\includegraphics[width=0.33\textwidth]{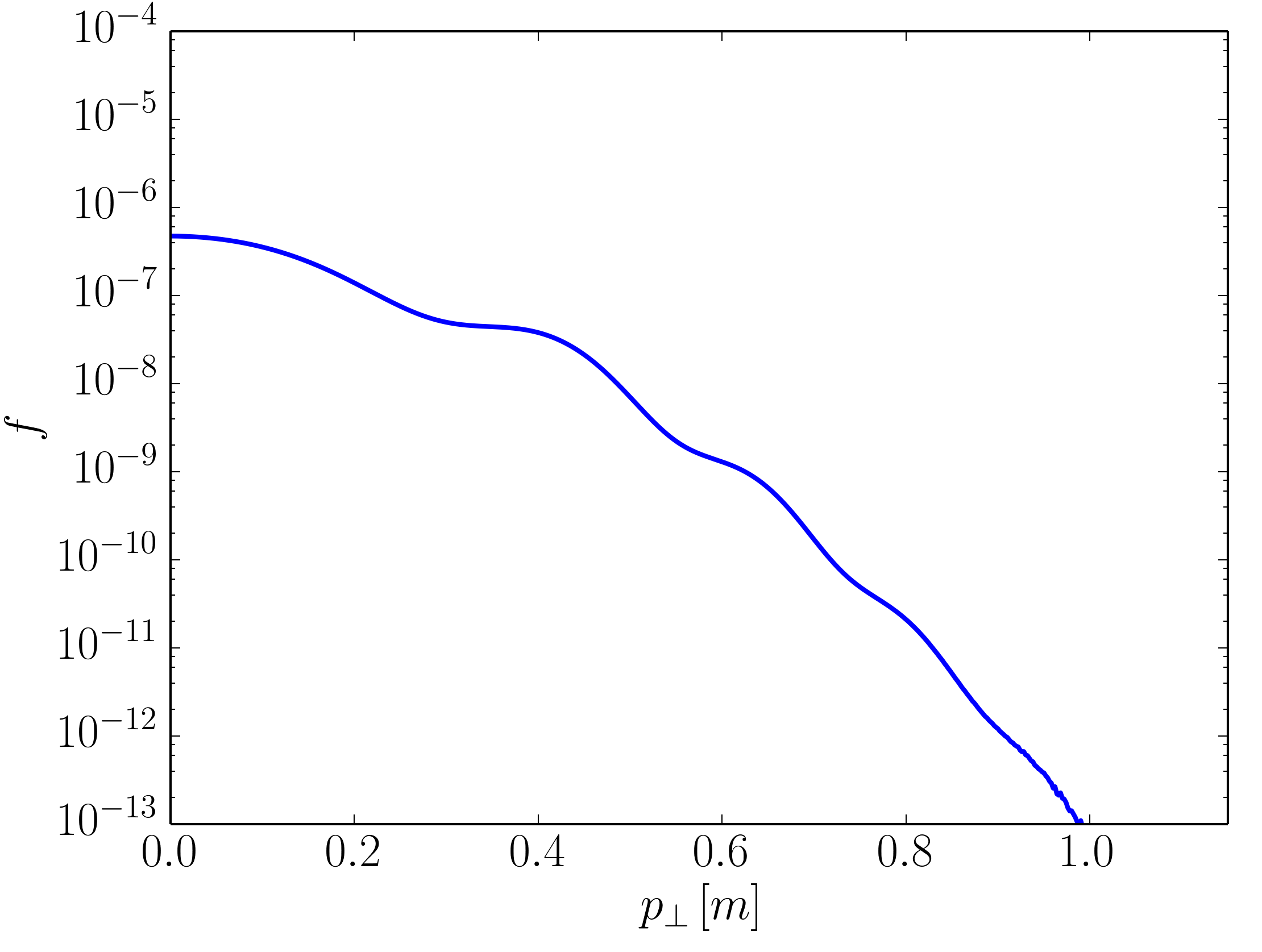}\hfill
\includegraphics[width=0.33\textwidth]{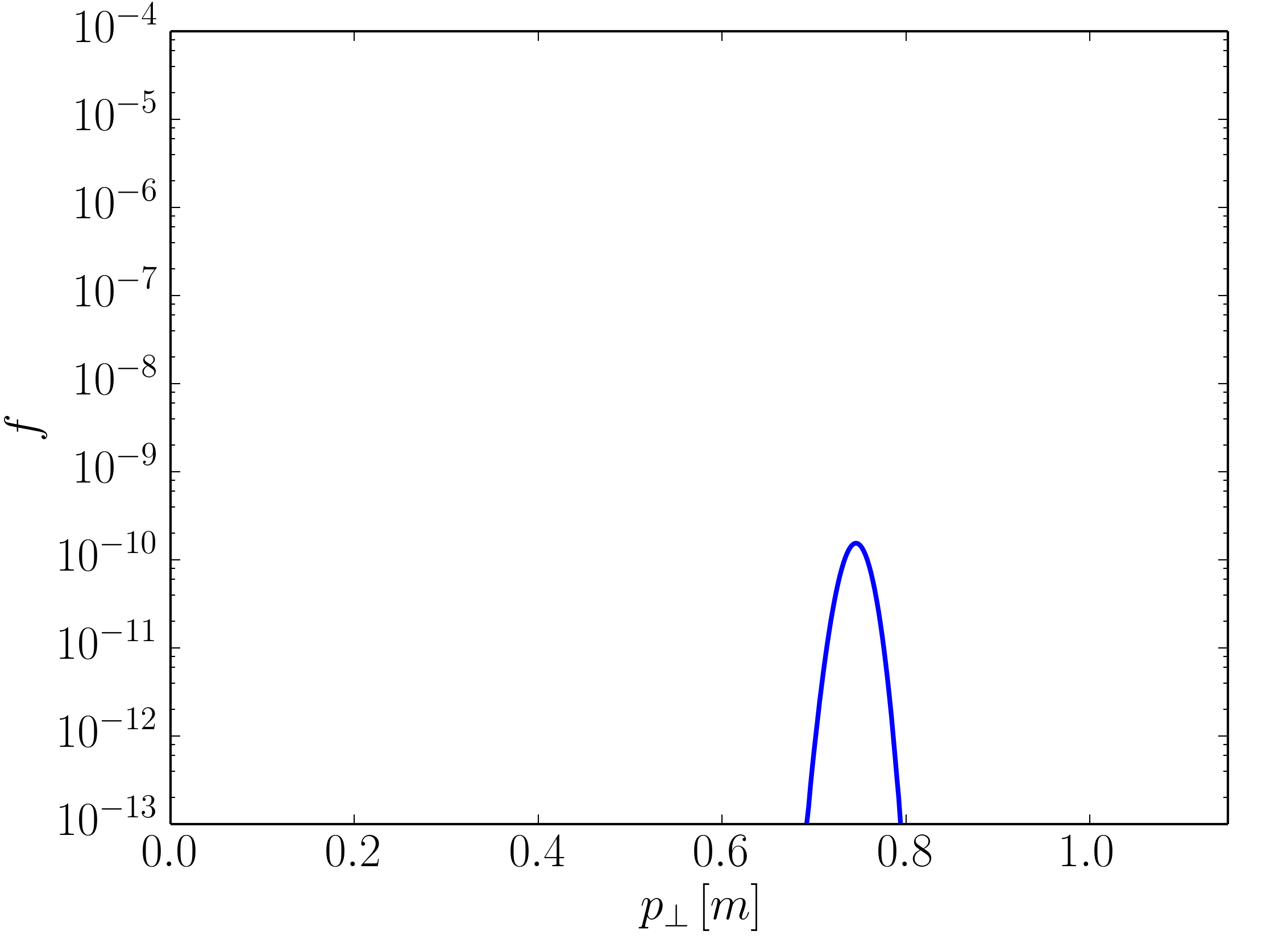}\hfill
\includegraphics[width=0.33\textwidth]{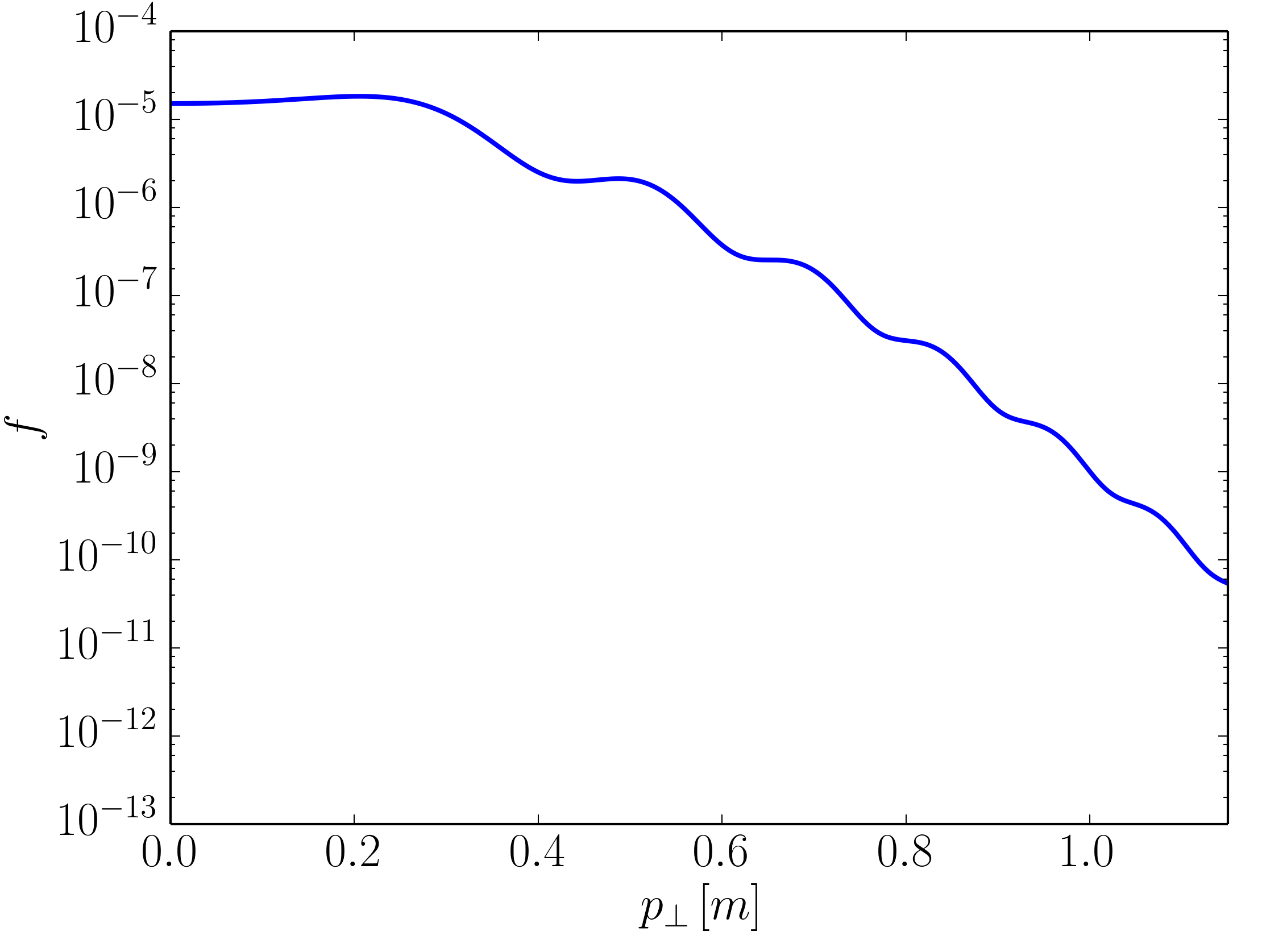}
\caption{Residual phase space distributions at $p_\perp = 0$ (upper row)
and $p_\parallel=0$ (lower row) for the fields displayed in Fig.~\ref{fig:1}. 
\label{fig:2}}
\end{figure*}

In Fig.~\ref{fig:2} we show the residual phase space distribution at $p_\perp = 0$ (upper row)
and $p_\parallel=0$ (lower row) for the fields displayed in Fig.~\ref{fig:1}. 
It is obvious that here the nonlinear parametric enhancement effect takes place.
The maximum values of the distribution function for the bifrequent pulse "1+2" 
are almost two orders larger than the corresponding values for the low-frequency pulse "1" 
and almost three orders of magnitude for the high-frequency pulse "2". 
In addition, the phase space occupancy for ``1+2'' is apparently strikingly larger.
Contrary to~\cite{Otto:2014ssa,Otto:2015gla},
one can hardly recognize a ``lifting'' of the $p_\parallel$ distribution for field ``1'' by ``2'':
The patterns are fairly different.
In so far, the enhancement patterns seem to be specific for the pulse shapes,
requiring individual investigation.

In contrast to oscillating fields with extended flat-top envelope \cite{Otto:2014ssa,Otto:2015gla},
the Gaussian envelopes in (\ref{field_G}) - (\ref{field_12GA}) with 
$\sigma = {\cal O} (5)$ do not allow for sharp resonance-like structures. 
Therefore, in this parameter domain, the density (\ref{dens_r}) is easier accessible.
Instead of $n$ we show in the following the dimensionless combination 
$N_{e^- e^+} = n / \omega^3$ which characterizes the number of pairs generated 
in a volume determined by the transverse size of the minimum focal spot attainable 
at the diffraction limit of field "1". 

\begin{figure*}
\includegraphics[width=0.5\textwidth]{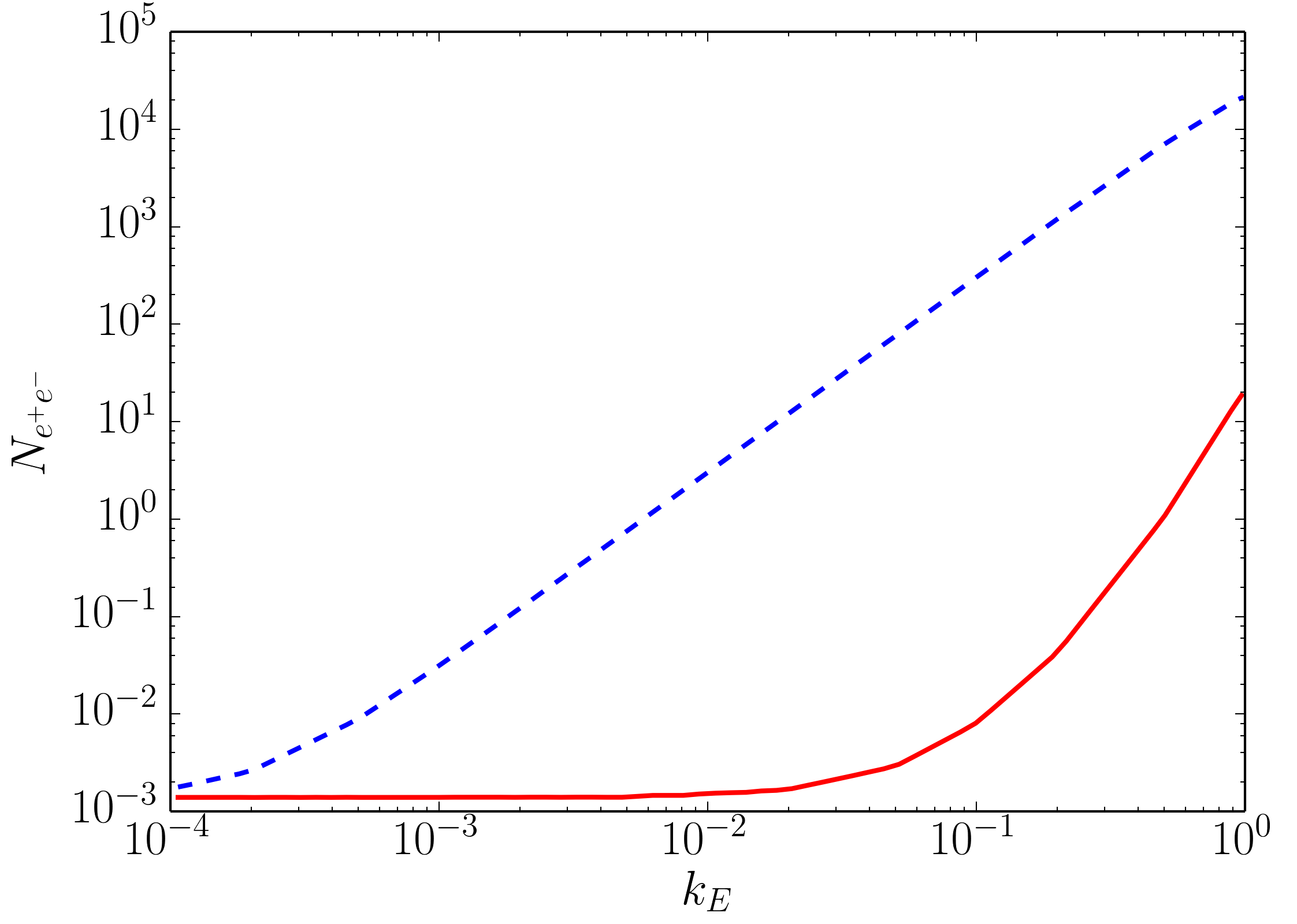}\hfill
\includegraphics[width=0.5\textwidth]{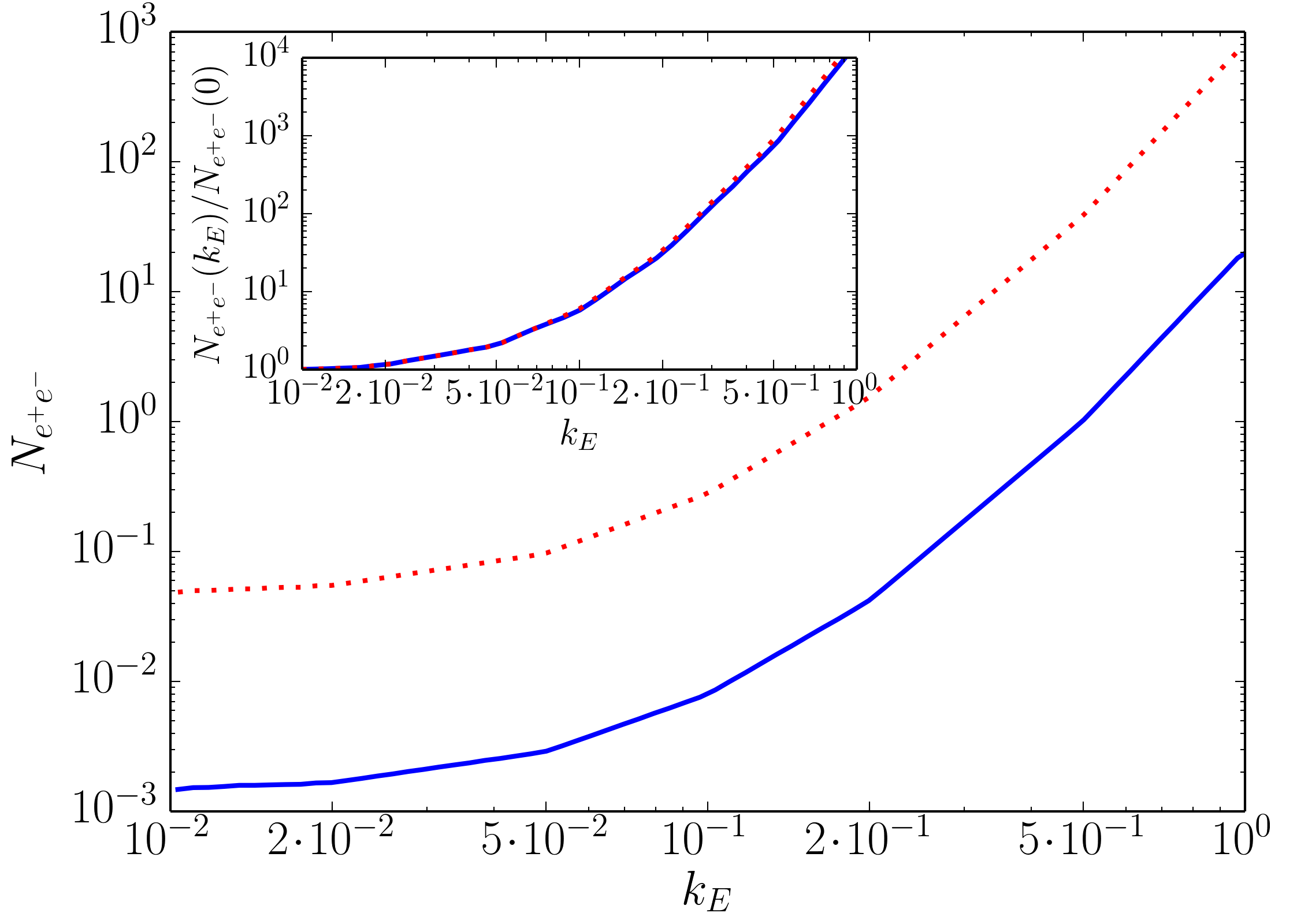}
\caption{The number of pairs produced in the pulse of Eqs.~(\ref{field_12G}) and (\ref{field_12GA}) 
with $E_0 = 0.2E_c$ and $\sigma = 5$.
Left: $\omega = 0.05~m$ and $k_\omega = 10$ (solid curve) and 
$k_\omega = 40$  (dashed curve).
Right: $\omega = 0.05~m$ (solid curve) and $\omega=0.02~m$ (dotted curve);
$k_{\omega}\omega = 0.5~m$ for both curves. 
The inset shows the ratio  $r = N_{e^-e^+} (k_E) / N_{e^-e^+}(0) = n_{1+2}/n_1$.
\label{fig:3}}
\end{figure*}

\begin{figure*}
\includegraphics[width=0.5\textwidth]{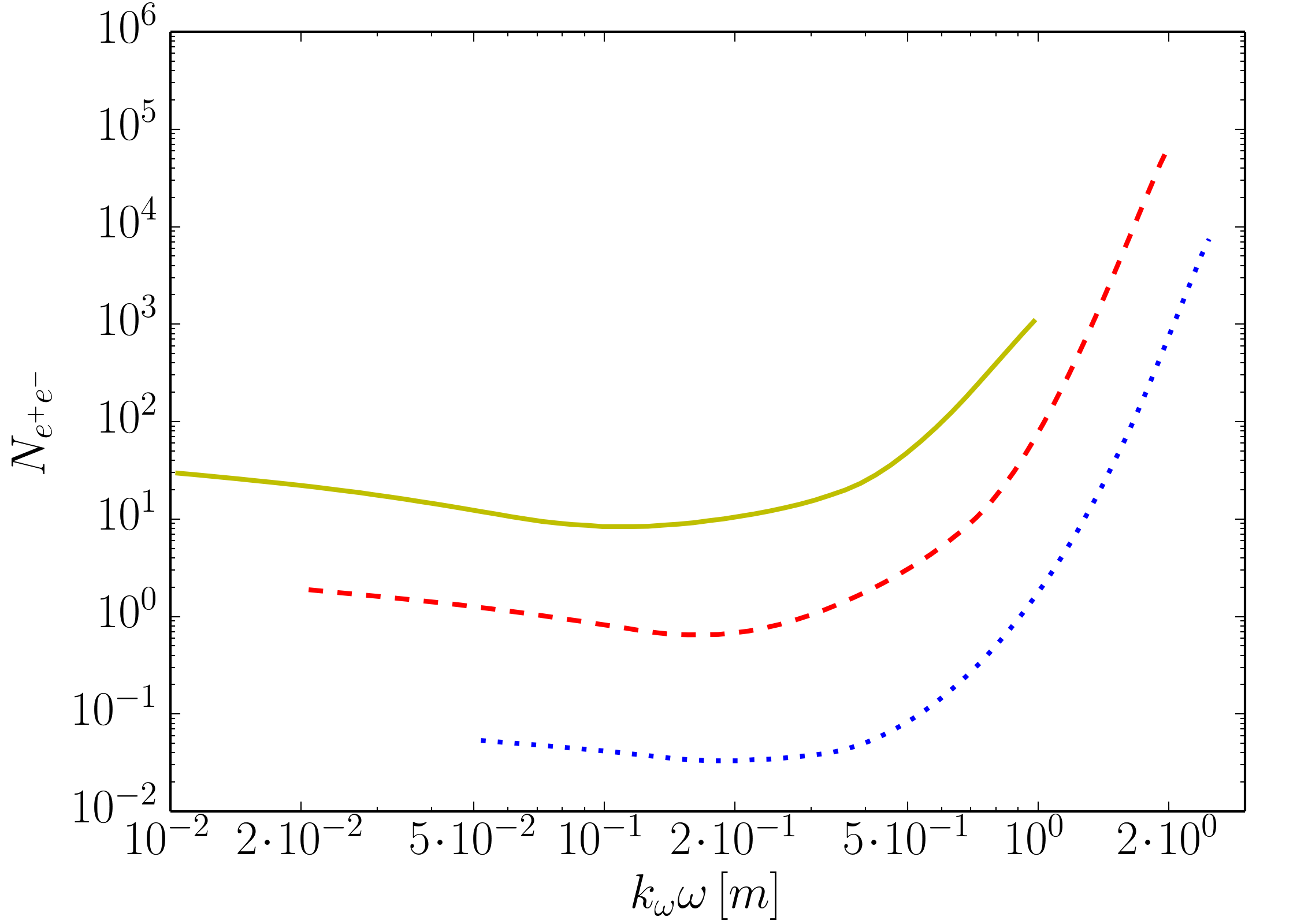}\hfill
\includegraphics[width=0.5\textwidth]{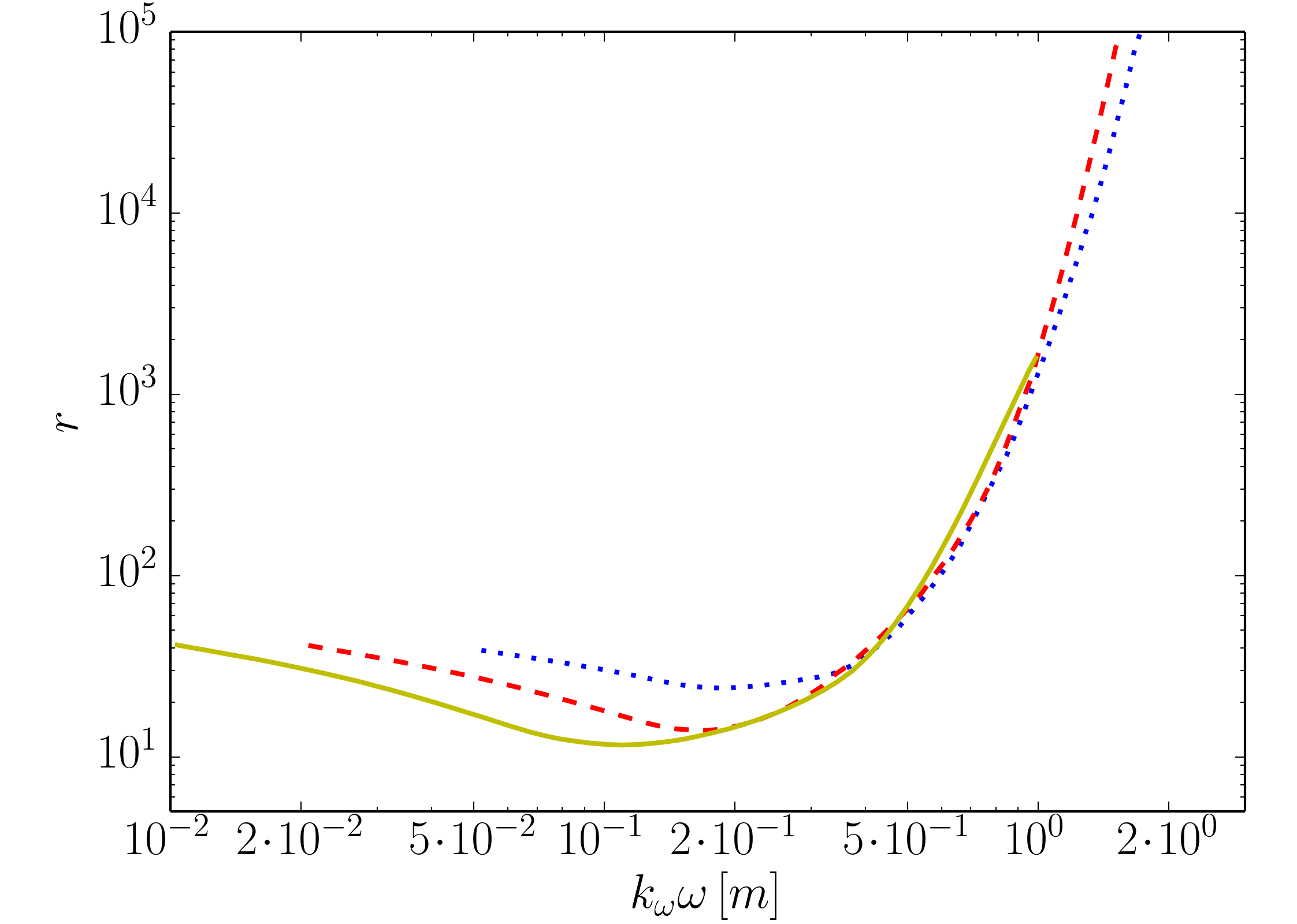}
\caption{Effectiveness of increasing the number of pairs produced for pulse type given by Eqs.~(\ref{field_12G}) and (\ref{field_12GA}) with $E_0 = 0.2~E_c$, $k_E = 0.25$ and $\sigma = 5$.
The number of pairs $N_{e^- e^+}$ (left) and the ratio $r$ (right) as a function of the weak-field frequency $k_\omega \omega$ for $\omega = 0.05~m$ (solid curve), $0.02~m$ (dashed curve) and $0.01~m$ (dotted curve).
\label{fig:4}}
\end{figure*}

Figure \ref{fig:3} shows the increase of the number of pairs created with increasing 
field strength $k_E E$ of the high-frequency pulse from small to large values of $k_E$. 
The left panel shows also a strong dependence of the effect 
on the frequency $k_\omega \omega$ of the second component of the field: 
At $k_\omega = 10$ 
the amplification effect becomes noticeable only for $k_E > 0.01$. 
For $k_{\omega} = 40$, an enhancement effect is seen already for $k_E > 0.0001$.
Such a behavior has been noted already in \cite{Otto:2015gla} for another special field model
and in \cite{Linder:2015vta} more generally: keeping fixed all other parameters, a certain
value of the field strength "2" is required to cause a noticeable amplification by
the assisting field.     
The right panel of Fig.~\ref{fig:3} shows that the effect is universal 
for different frequencies $\omega$ of the strong field "1". The effect depends weakly on 
$\omega$ at fixed high-frequency $k_\omega \omega$. 
In the inset of that panel, we show the ratio $r = N_{e^-e^+} (k_E) / N_{e^-e^+}(0) = n_{1+2}/n_1$ 
as a function of $k_E$ to quantify the amplification effect.
In particular, at $k_e \to 1$, the enhancement due to the assisting field becomes 
enormously large.

The dependence of the amplification effect on the frequency $k_\omega \omega$
of the weak, high-frequency field component 
is presented in Fig.~\ref{fig:4}. In the left panel, the dependence of the number of created pairs is presented for three values of the strong-field frequency $\omega$. 
At the same time, the frequency range of the second field component runs 
in each case over a range from values of the frequency $\omega$,
i.e.\ $k_{\omega} = 1$, up to $k_{\omega} \omega = 2~m$. 
The limiting case of equality of the first and the second frequency components is
equivalent to an increase of the field amplitude of the first component by 
the coefficient $1 + k_E$ and corresponds to the field defined by Eqs.~(\ref{field_G}) and (\ref{field_GA}) with  $E_0 \to E_0 ( 1+ k_E)$.  
The right panel of Fig.~\ref{fig:4} shows the dependence on the increase of the number of pairs created by the second component. 
For all three pulse frequencies $\omega$, the results are almost identical, 
and at high frequencies $k_{\omega} \omega \approx m$ fairly large.
The right panel of  Fig.~\ref{fig:4} shows that, for relatively low values of  the frequency 
$k_{\omega} \omega < 0.5~m$, the amplification effect is practically independent of 
the frequency and is equivalent to the corresponding increase of the amplitude of one field by
$E_0 \to E_0 (1 + k_E)$ at $k_\omega = 1$. There is a slight decrease of $r$ in the range
$1 < k_{\omega} \lesssim 10$. 
A likely explanation is the interference of two pulses with similar parameters.

\begin{figure*}
\includegraphics[width=0.33\textwidth]{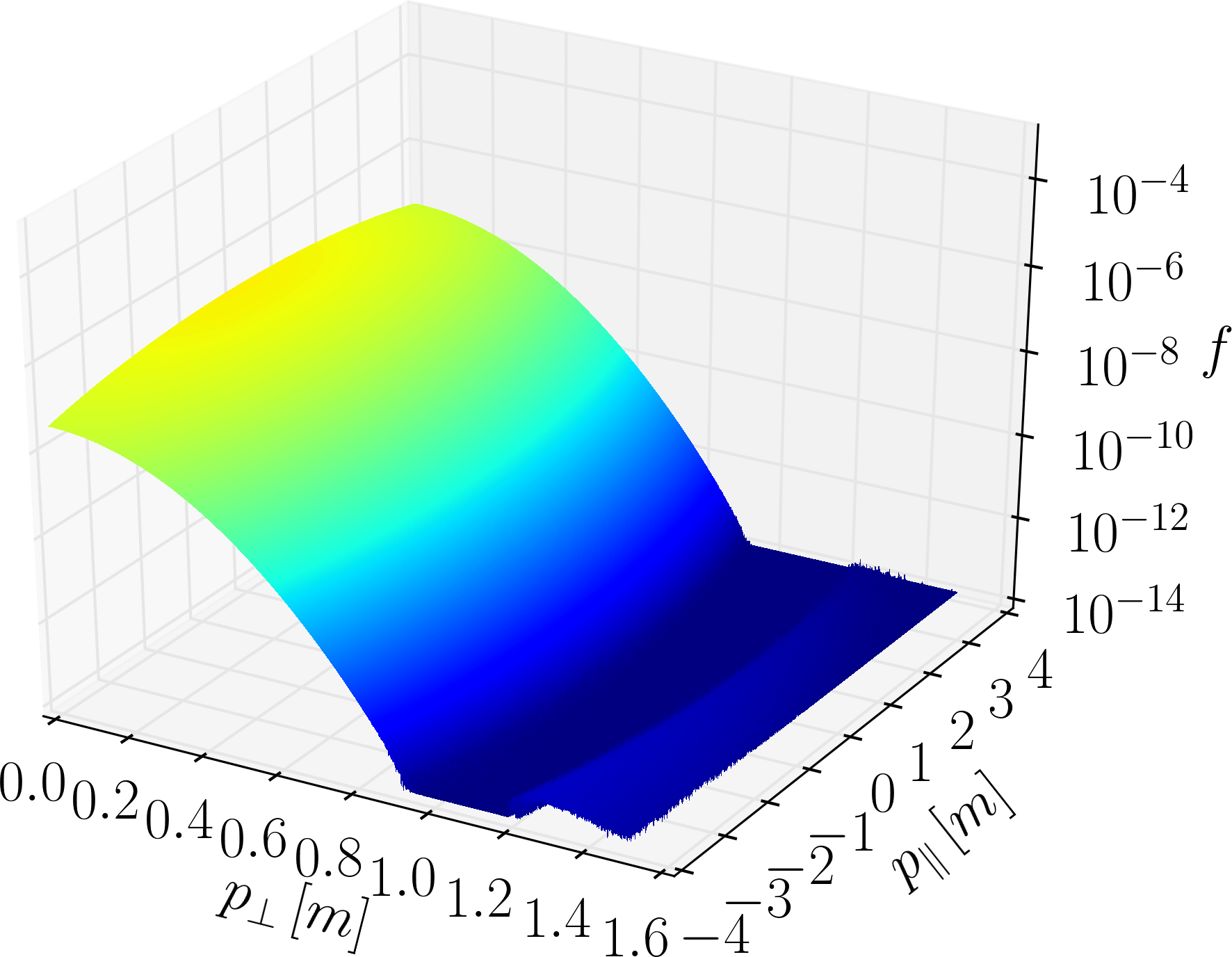}\hfill
\includegraphics[width=0.33\textwidth]{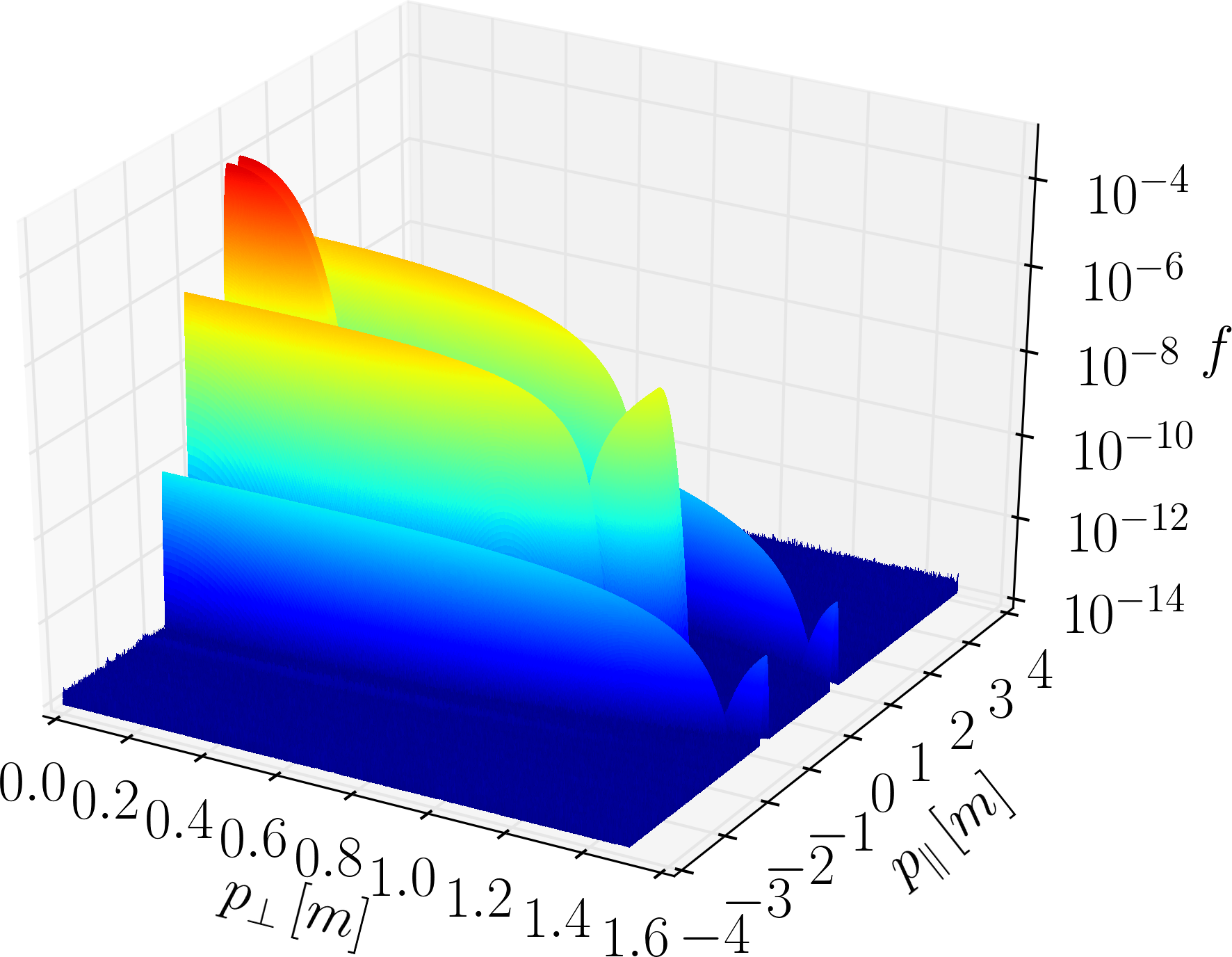}\hfill
\includegraphics[width=0.33\textwidth]{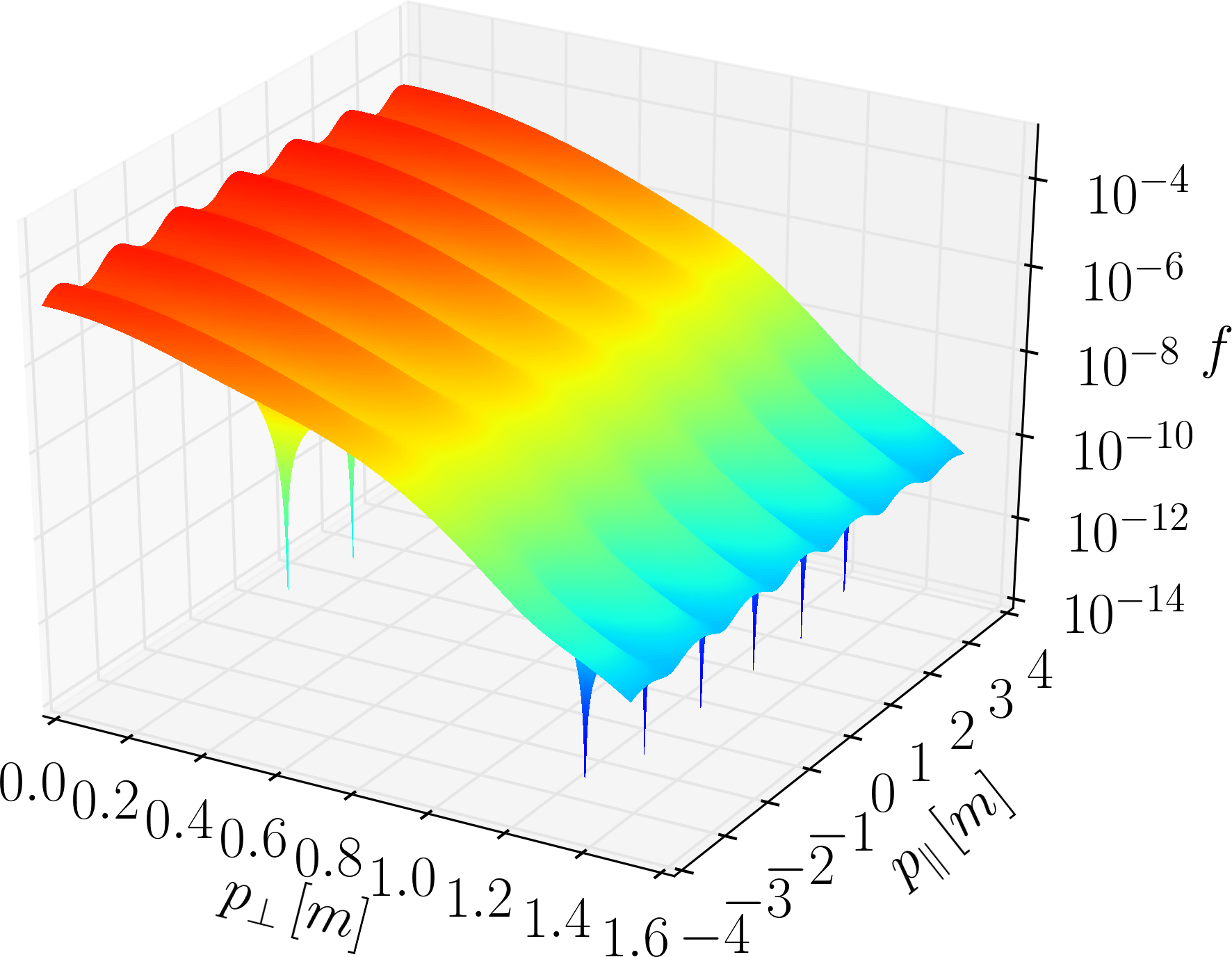}
\caption{The phase distribution for the pulse (\ref{new_pulse}) with a 
Gaussian envelope with $E_0=0.2~E_c$ and $\sigma = 50$.
Left: $k_E = 0$,
middle: only for the second term in  (\ref{new_pulse}) with $k_E = 0.25$, $\omega_2 = m$, 
right: $k_E = 0.25$, $\omega_2 = m$. 
\label{fig:5}} 
\end{figure*}

It should be stressed that the pair production in the
multi-photon regime is becomes very efficient
and depends less on the field strength. 
To illustrate that point let us consider
the pulse model (\ref{new_pulse}) with a Gaussian envelope and 
$E_0 = 0.2~E_c$ and $\omega_2 = m$.
For $k_E = 0$, i.e.\ only the first term in  (\ref{new_pulse}), the phase space
distribution is smooth (see the left panel of Fig.~\ref{fig:5}), 
in contrast to the distribution shown in the left panel of Fig.~2. 
For larger values of $\sigma$, the distribution approaches that of the Schwinger process, which is
flat in $p_\parallel$ direction and Gaussian shaped in $p_\perp$ direction.
In the displayed momentum range, one pronounced multi-photon peak is visible when considering
the second term in (\ref{new_pulse}) alone, see the middle panel of Fig.~\ref{fig:5};
it is accompanied by much lower side-ridges in $p_\perp$ direction
(the cross section at $p_\perp=0$ looks similar to the middle panel of Fig.~\ref{fig:2}, of course).
Its peak value is much higher than the maximum seen in the left panel, even the field
strength is less. That is the efficiency of the multi-photon process.
The complete pulse  (\ref{new_pulse}) gives rise to the phase distribution exhibited 
in the right panel. The enhancement relative to the left panel is obvious, but the net effect
falls short in comparison to the middle panel, when comparing the maxima of $f$.
In the example at hand, the action of field ``2'' looks more like a ``lifting''
of the distribution emerging from ``1'', albeit without the ripples.
While the ratio $r = n_{1+2} / n_1$ rises strongly for $\omega_2 \to m$
(as seen in the right panels of Figs.~\ref{fig:3} and~\ref{fig:4} for another pulse),
the net efficiency $n_{1+2} / (n_1 + n_2)$ acquires a maximum which can be much
larger than unity, but drops ultimately to unity upon enlarging further $\omega_2$
as emphasized in \cite{Orthaber:2011cm}.
It is the distinct phase space distribution
which becomes important to discriminate the impact of the field components.

\section{Discussion \label{sect:disco}}

Our investigation was originally motivated by the availability of XFELs
($E_{\rm XFEL} \sim 10^{-5} E_c$, $\omega_{\rm XFEL}  \sim 5 - 50$ keV,
cf.\ figure 1 in \cite{Ringwald:2001ib} and \cite{Otto:2015gla}) and PW laser systems
($E_{\rm PW} \sim 10^{-3} E_c$, $\omega_{\rm PW}  \sim1 -3$ eV, cf.\
\cite{eli,hiper,Dunne:2014qda}). 
These installations, when being combined with each other
(as envisaged in the HIBEF project \cite{hibef} for instance, 
or available already at LCLS \cite{lcls}), in principle, would be
characterized by $k_\omega > 10^3$ and $k_E \sim 10^{-2}$. 
Moreover, pulse lengths of sub-attosecond duration would correspond to $m \tau \sim 10^2$. 
Clearly, these values are fairly distinct from those we have considered above. 
Thus, our present considerations do
not directly apply to situations which can be expected to be exploited for
experimental investigations towards the assisted dynamical Schwinger effect. 
In so far, our work is an exploratory supplement to studies searching for promising designs
with discovery potential w.r.t.\ genuinely non-perturbative mechanisms of particle production. 
Without strikingly new ideas on avenues to the experimental verification
of the Schwinger effect in freely propagating fields (in contrast to the nuclear
Coulomb field), the many details understood by now call for significantly higher
fields and/or large photon frequencies. 
Nevertheless, the facets of the Schwinger effect remain challenging, in particular due to their 
relation to many other fields as quoted in the Introduction.         

\section{Summary \label{sect:sum}}

When two pulses with different frequencies and different field strengths 
(the latter ones being high enough to be not only a too small fraction of $E_c$) 
one can talk about two mechanisms for the increase of the pair production. 
If the frequencies of the two components are close 
(in the extreme case, we even can assume they are the same) 
and are small compared to the energy required for 
multi-photon pair creation, the nature of the  increment of residual pairs is directly related to the highly non-linear dependence of the effect on the field  strength in the vicinity of $E_c$. Alternatively, when one of the frequencies is not high and the second one is approaching the threshold of pair production by single photons we can talk about changing the properties of the vacuum for the high-energy photons. In this case, we can expect to more effectively promote the process of pair production and consider this process as pair production by a short-wavelength component
catalyzing the low-frequency component.

In the present study we demonstrate that the increase of the rate of $e^-e^+$ production by combining a strong low-frequency field and a weak high-frequency field is a universal phenomenon and manifests itself in a certain range of parameters of the high-frequency field. 
Our results have been obtained within a non-perturbative framework. 
The shape of the electric field pulse is realistic and reproduces to some extent the characteristics of field pulses in experimental setups. 
The presented approach allows on the one hand to optimize the parameters for practical implementations of the dynamical Schwinger effect. 
On the other hand, by choosing parameters of the field model that characterise the actual experiment it allows to accurately estimate the number of residual pairs and their characteristics.

\subsection*{Acknowledgments}
The authors acknowledge fruitful discussions with R.~Sauerbrey, T.~E.~Cowan
and D.~Seipt within the HIBEF project.
A.D.P. received support from the University of Wroclaw under internal number 
2467/M/IFT/14 for his visit  at the Institute for Theoretical Physics.
D.B. and L.J. have been supported by Narodowe Centrum Nauki under grant number
UMO-2014/15/B/ST2/03752.

\end{document}